\documentclass{article}
\usepackage{spconf,amsmath,graphicx,booktabs, amssymb}
\usepackage[subrefformat=parens]{subcaption}

\usepackage{hyperref}
\usepackage{listings}
\usepackage{url}
\usepackage{soul}
\usepackage{algorithm}
\usepackage{algorithmic}
\usepackage{caption}
\usepackage{booktabs}
\usepackage{color}
\usepackage{relsize}

\def\x{{\mathbf x}}

\usepackage{multirow}
\usepackage{rotating}

\makeatletter
\def\Hline{%
\noalign{\ifnum0=`}\fi\hrule \@height 1pt \futurelet
\reserved@a\@xhline}
\makeatother

\usepackage{pifont} 

\definecolor{mygreen}{rgb}{0.032, 0.6392, 0.2039}
\newcommand{\cmark}{\textcolor{mygreen}{{\boldmath $\bigcirc$}}}%
\newcommand{\xmark}{\textcolor{red}{\ding{55}}}%

\title{Effort-free Automated Skeletal Abnormality Detection\\ of Rat Fetuses on Whole-body Micro-CT Scans}

\name{Akihiro Fukuda \quad Changhee Han \quad Kazumi Hakamada}

\address{LPIXEL Inc., Japan}

\begin{document}

\maketitle
\begin{abstract}
Machine Learning-based fast and quantitative automated screening plays a key role in analyzing human bones on Computed Tomography (CT) scans. However, despite the requirement in drug safety assessment, such research is rare on animal fetus micro-CT scans due to its laborious data collection and annotation. Therefore, we propose various bone feature engineering techniques to thoroughly automate the skeletal localization/labeling/abnormality detection of rat fetuses on whole-body micro-CT scans with minimum effort. Despite limited training data of $49$ fetuses, in skeletal labeling and abnormality detection, we achieve accuracy of $0.900$ and $0.810$, respectively.
\end{abstract}


\section{Introduction}\label{sec:intro}
\vspace{-1mm}
Manufacturing (bio)pharmaceuticals requires strict Developmental And Reproductive Toxicology (DART) studies on animal fetuses to investigate drug-induced birth defects~\cite{winkelmann2009high,wise2010micro,french2010use}. Although manual inspection with Alizarin Red staining is current practice for detecting skeletal abnormalities, it is a long, laborious, and non-quantitative procedure~\cite{dogdas2015characterization}. In this regard, Machine Learning can enable fast and quantitative automated screening; it has shown great promise in human skeletal 
localization~\cite{noguchi2020bone}, labeling~\cite{belal2019deep}, and abnormality detection~\cite{li201918f} on CT scans.

Recently micro-CT comes into use in this field~\cite{solomon2018use}. However, few researchers applied Machine Learning on animal fetus micro-CT scans due to its labor-intensive nature---one needs to collect/annotate significant number of whole-body micro-CT scans. As the only work, Chen \textit{et al.} localized and labeled cervical vertebral bones of rabbit fetuses~\cite{chen2019localization}. However, researchers have tackled none of (\textit{i}) skeletal abnormality detection, (\textit{ii}) 
whole-body bones, and (\textit{iii}) rat fetuses.

\begin{figure}[t!]
\includegraphics[width=8.5cm]{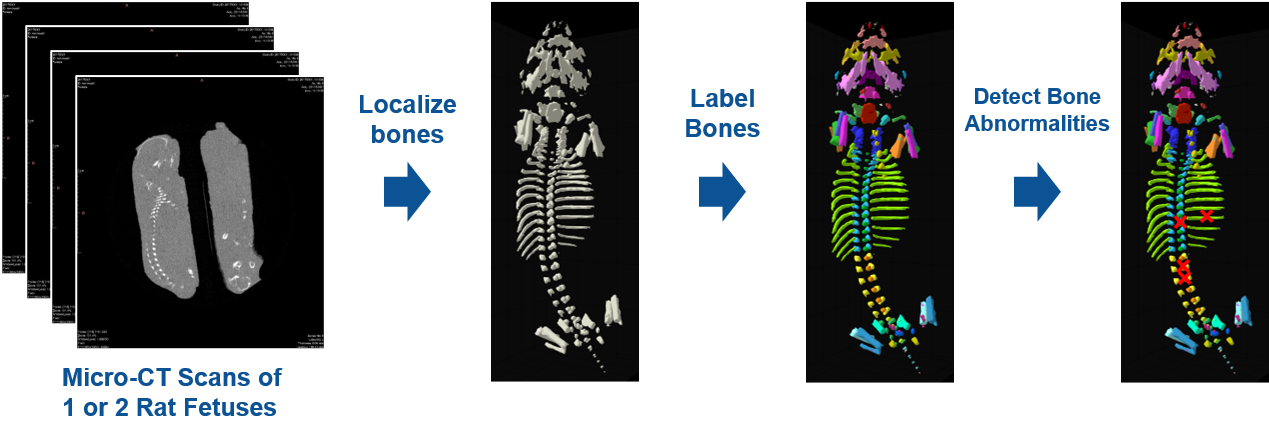} 
\caption{Overview of our approach for effort-free  whole-body  skeletal  abnormality  detection.}\label{fig:whole-flow}
\vspace{-3mm}
\end{figure}

How can we thoroughly automate the skeletal localization/labeling/abnormality detection of rat fetuses on whole-body micro-CT scans with minimum effort? As shown in Fig. \ref{fig:whole-flow}, we propose various bone feature engineering techniques for (\textit{i}) unsupervised localization to reduce human annotation effort; (\textit{ii}) supervised labeling adopting Body Axis Correction (BAC)-based registration to alleviate the need for rigorous subject positioning in micro-CT; (\textit{iii}) simple rule-based abnormality detection. 





\vspace{1mm}
\textbf{Contributions.} Our main contributions are as follows:\vspace{-2mm}
\begin{itemize}
\item \textbf{Automated Micro-CT Analysis:} This quantitative study firstly analyzes whole-body micro-CT scans of rat fetuses using Machine Learning.\vspace{-2mm}
\item \textbf{Effort-free Skeletal Labeling:} Our architecture adopting BAC, feature registration, accurately classifies bones into $40$ types under limited micro-CT training data without rigorous subject positioning.\vspace{-2mm}
\item \textbf{Effort-free Skeletal Abnormality Detection:} Our three simple rules reliably detect skeletal abnormalities towards fast and automated screening (e.g., drug safety assessment) on micro-CT scans.
\end{itemize}

\begin{figure*}[t]
 \centering
\includegraphics[width=15cm]{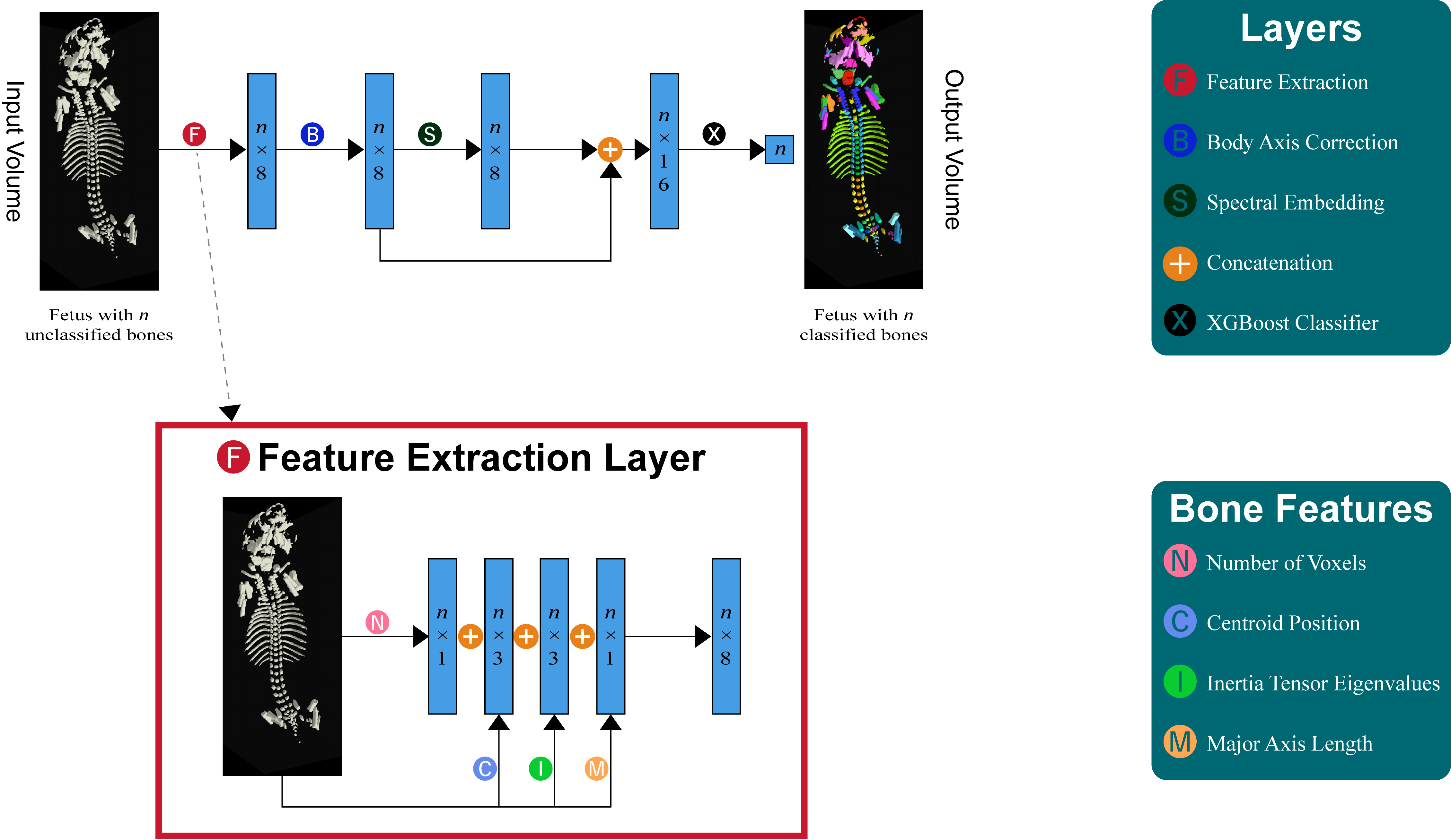} 
\vspace{-2mm}
\caption{Architecture for effort-free skeletal labeling: it classifies $n$ bones (each fetus contains a different number of bones) into $40$ types, respectively. BAC enable to perform effective registration without rigorous subject positioning in micro-CT.}\label{fig:cls-flow}
\vspace{-3mm}
\end{figure*}

\vspace{-3mm}
\section{Materials and Methods}\label{sec:method}
\vspace{-1mm}

\subsection{Rat Fetus Dataset}\label{ssec:data}
We use a dataset of $86$ whole-body micro-CT scans containing $166$ rat fetuses---each scan contains up to 2 fetuses. The dataset was collected by Astellas Pharma Inc. and currently not publicly available for commercial restrictions. The scans have $1024 \times 1024 \times 317$ average resolution with $0.06$ mm isotropic voxel size. 3D median filtering~\cite{viero1994three} mitigates speckle noise in the scans. 
By clicking pre-localized 3D bone meshes (more details of the skeletal localization algorithm are given later), expert DART scientists categorize the bones into $40$ types, including nasal bone; cervical vertebral body; thoracic vertebral body; lumbar vertebral body; rib; tibia.


Based on skeletal malformation observed in Alizarin Red staining, we divide all $166$ fetuses randomly as follows:
\begin{itemize}
\item \textbf{Training set}: Normal ($12$), Abnormal ($37$);
\vspace{-2mm}
\item \textbf{Validation set}: Normal ($3$), Abnormal ($9$);
\vspace{-2mm}
\item \textbf{Test set}: Normal ($34$), Abnormal ($71$).
\end{itemize}

Since drug safety assessment demands a large sample size~\cite{watkins2011drug}, the test set contains more fetuses than the training/validation sets. Such a problem definition (i.e., training with limited data) is essential since it is laborious to collect/annotate normal/abnormal whole-body micro-CT scans.

\vspace{-2mm}
\subsection{Effort-free Skeletal Abnormality Detection 
Approach}\label{ssec:analysis}
\subsubsection{Skeletal Localization}\label{sssec:localize}
To reduce experts' annotation efforts, we automate a skeletal localization process in an unsupervised manner. We first extract each fetus' bones from micro-CT scans composed of up to two rat fetuses $via$ watershed thresholding~\cite{shojaii2005automatic}; then, we convert the corresponding bones into 3D meshes $via$ the marching cubes algorithm~\cite{lewiner2003efficient}.\vspace{2mm}

\noindent \textbf{Watershed Thresholding Implementation Details}
Based on Hounsfield Units (HUs), we first classify input voxels into the background (HUs below 430), border (HUs between 430 and 580), and bone (HUs above 580). Then, watershed thresholding classifies the remaining border voxels.
\vspace{-3mm}



\begin{table*}[t!]
\centering
\caption{Skeletal labeling results for ablations of included features: $\x$, $\mathcal{B}$, and $\mathcal{S}$ represent bone features, BAC, and spectral embedding, respectively. Sensitivity and precision are calculated using macro averages of all classes.}\label{tab:res-cls}
\begin{tabular}{cccccccc}
\Hline\noalign{\smallskip}
& \multicolumn{4}{c}{\textit{Included Features}} & & & \\
 \bfseries Model & \bfseries $\x(i)$ & \bfseries $\mathcal{S}(\x(i))$ & \bfseries $\mathcal{B}(\x(i))$ & \bfseries $\mathcal{S}(\mathcal{B}(\x(i)))$ &\bfseries Accuracy[\%] & \bfseries Sensitivity[\%] & \bfseries Precision[\%]
 \\\noalign{\smallskip}\hline\noalign{\smallskip}
 SVM &  \xmark & \xmark & \cmark & \cmark & 73.9 & 64.3 & 63.2\\
 \noalign{\smallskip}\hline\noalign{\smallskip}
 LightGBM  &  \cmark & \cmark & \xmark & \xmark  & 73.0 & 64.7 & 63.7\\
 LightGBM &  \xmark & \xmark & \cmark & \xmark  & 88.3 & 83.7 & 83.8\\
 LightGBM &  \xmark & \xmark & \xmark & \cmark  & 81.3 & 76.2 & 76.9\\
 LightGBM &  \xmark & \xmark & \cmark & \cmark  & 89.5 & 86.1 & 86.5\\
 \noalign{\smallskip}\hline\noalign{\smallskip}
 XGBoost  &  \cmark & \cmark & \xmark & \xmark  & 73.8 & 65.9 & 65.0\\
 XGBoost &  \xmark & \xmark & \cmark & \xmark  & 87.4 & 82.6 & 82.5\\
 XGBoost &  \xmark & \xmark & \xmark & \cmark  & 81.7 & 76.6 & 77.4\\
 XGBoost &  \xmark & \xmark & \cmark & \cmark  & \textbf{90.0} & \textbf{86.7} & \textbf{86.8}\\
\noalign{\smallskip}\Hline\noalign{\smallskip}
\end{tabular}
\vspace{-3mm}
\end{table*}

\subsubsection{Skeletal Labeling}\label{sssec:classify}
We propose a novel architecture for accurate skeletal labeling with limited training data (Fig. \ref{fig:cls-flow}). It automatically labels bones using various bone-related feature engineering techniques on micro-CT scans before classification with a classical machine learning method. We extract the following bone features $\x(i)\in \mathbb{R}^{n \times 8}$ for a given pre-localized 3D bone volume $i$ with $n$ unclassified fetus bones:
\vspace{-2.5mm}

\begin{equation}
\begin{aligned}
\x(i) &= \mathcal{N}(i) + \mathcal{C}(i) + \mathcal{I}(i) + \mathcal{M}(i),
\end{aligned}
\end{equation}

\noindent where $\mathcal{N}(i)\in \mathbb{R}^{n \times 1}$ denotes each bone's number of voxels, $\mathcal{C}(i)\in \mathbb{R}^{n \times 3}$ is each bone's centroid position, $\mathcal{I}(i)\in \mathbb{R}^{n \times 3}$ indicates each bone's inertia tensor eigenvalues, and $\mathcal{M}(i)\in \mathbb{R}^{n \times 1}$ represents each bone's major axis length. We use those features to capture bone size, shape, and structure.

Then, as an effective bone feature registration technique, we apply Body Axis Correction (BAC) $\mathcal{B}(\x(i)) (\mathcal{B} : \mathbb{R}^{n \times 8} \rightarrow \mathbb{R}^{n \times 8})$ to improve classification without requiring rigorous subject positioning in micro-CT. The BAC exploits two Principal Component Analysis (PCA~\cite{wold1987principal}) steps: (\textit{i}) the rough alignment of body axes using whole $\langle \mathcal{N}(i),\mathcal{C}(i) \rangle$, followed by correcting the head/tail direction with primary axis flipping; (\textit{ii})  the further alignment of the spine direction using $\langle \mathcal{N}(i),\mathcal{C}(i) \rangle$ only at the negative direction of the primary axis. We also apply spectral embedding~\cite{luo2003spectral} $\mathcal{S}(\mathcal{B}(\x(i))) (\mathcal{S} : \mathbb{R}^{n \times 8} \rightarrow \mathbb{R}^{n \times 8})$ to emphasize each bone's similarity, connectivity, and global structure. Since the spectral embedding via Laplacian matrix decomposition maps neighboring points on the manifold to neighboring points again while preserving local distances, it differently represents the BAC-extracted geometrically close bone features for mutual complementation. Finally, we concatenate the features $\mathcal{B}(\x(i))$ and $\mathcal{S}(\mathcal{B}(\x(i)))$ to capture both linear/non-linear relations and classify $n$ bones into $40$ types using an eXtreme Gradient Boosting (XGBoost~\cite{chen2015xgboost}) classifier $\mathcal{
X}(\mathcal{X} : \mathbb{R}^{n \times 16} \rightarrow \mathbb{R}^{n})$. We also compare the XGBoost against other well-defined classifiers: Support Vector Machine (SVM~\cite{noble2006support}) and Light Gradient Boosting Machine (LightGBM~\cite{ke2017lightgbm})---we do not implement Deep Learning since it generally requires large-scale training data.\vspace{2mm}

\noindent \textbf{Classifier Implementation Details}
~
For training classifiers, we use hyper-parameters as follows: XGBoost adopts 718 estimators with learning rate 0.0394 and maximum depth 8; LightGBM uses 143 estimators with learning rate 0.1037, maximum depth 8, and 171 leaves; SVM exploits RBF kernel with C of 39.6 and gamma of 0.04.
\vspace{-2mm}
\subsubsection{Skeletal Abnormality Detection}\label{sssec:abnormality}
\begin{figure}[t!]
 \begin{minipage}[b]{0.48\linewidth}
\centering   \includegraphics[width=\linewidth]{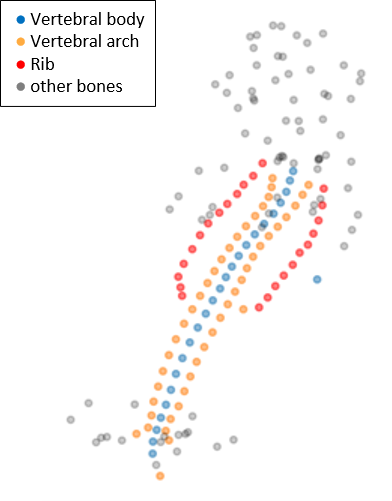}
  \subcaption{}
 \end{minipage}
 \begin{minipage}[b]{0.48\linewidth}
\centering   \includegraphics[width=\linewidth]{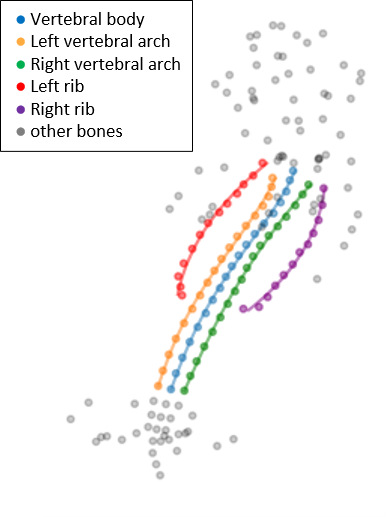}
  \subcaption{}
 \end{minipage}
  \vspace{-2mm}
\caption{An example of skeletal labeling results (a) without and (b) with curve fitting-based skeletal relabeling: each point represents a vertebral body, vertebral arch, or rib.}\label{fig:postprocess}
\vspace{-2mm}
\end{figure}

We focus on detecting skeletal abnormalities in vertebral bodies, vertebral arches, and ribs, where the skeletal abnormality is frequently observed. To improve skeletal labeling among those bones, we first apply curve fitting-based skeletal relabeling (Fig. \ref{fig:postprocess}). Since the vertebral bodies, vertebral arches, and ribs should line up along the curves respectively, we fit the curves to bone centroids and fix bone labels by following their nearest curves. Then, as our preliminary rule-based analysis, we 
detect skeletal abnormalities (i.e., binary classification on bone volumes) with the following three simple rules based on DART scientists’ manual inspection:

\begin{itemize}
\vspace{-2mm}
\item \textbf{Rule 1}: Neighboring vertebral bodies should possess a similar number of voxels ($\Delta < 20\%$);
\vspace{-2mm}
\item \textbf{Rule 2}: The total number of thoracic/lumbar vertebral bodies should be 19 or more;
\vspace{-2mm}
\item \textbf{Rule 3}: The most caudal rib and its neighboring rib should possess similar major axis length ($\Delta < 50\%$). 
\end{itemize}

\vspace{-2mm}
\noindent \textbf{Skeletal Relabeling Implementation Details}
Using the weighted least squares method, we fit quadratic curves for left/right ribs and quartic curves for vertebral bodies and left/right vertebral arches, respectively. (\textit{i}) Before curve fitting, we exclude points whose centroids are located at the caudal side of a line connecting head-side ends of bilateral ileum. (\textit{ii}) During curve fitting, each point has a weight according to reciprocal distances between its centroid and centroids of neighboring points within the same label. (\textit{iii}) After curve fitting, we exclude points whose distances to the corresponding curves are larger than the standard deviation of distances among all points within the same label. Before curve fitting, we exclude points whose distance to the corresponding curves is larger than the standard deviation of distances among all points within the same label.

\vspace{-2mm}
\section{Results}\label{sec:results}
\vspace{-2mm}

\subsection{Skeletal Labeling}\label{ssec:res-cls}
\begin{figure}[b!]
 \begin{minipage}[b]{0.48\linewidth}
\centering   \includegraphics[width=\linewidth]{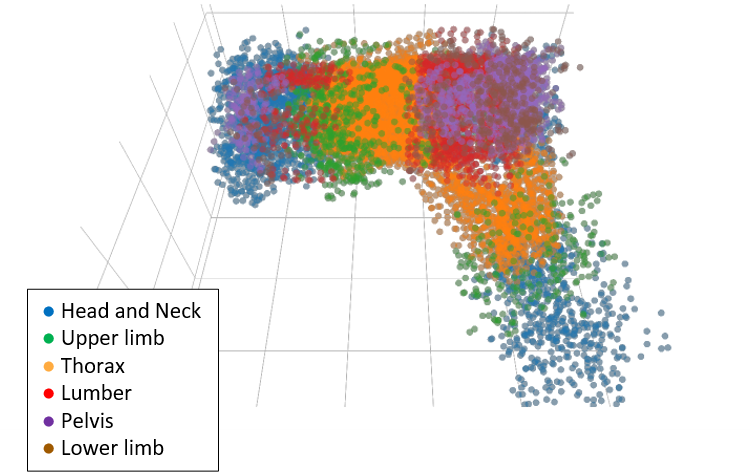}
  \subcaption{}
 \end{minipage}
 \begin{minipage}[b]{0.48\linewidth}
\centering   \includegraphics[width=\linewidth]{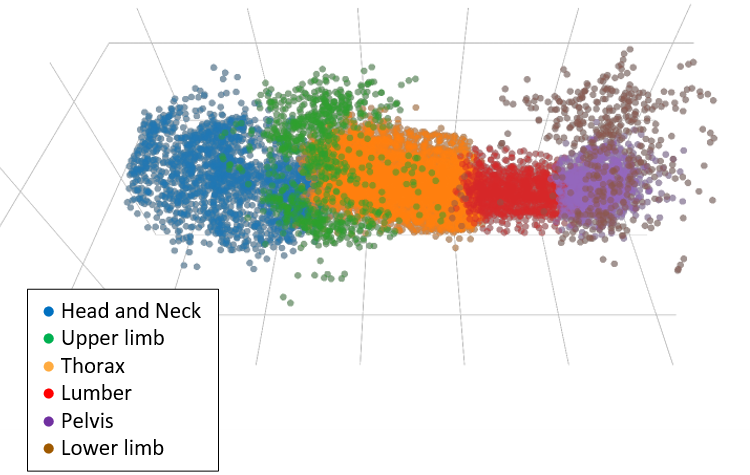}
  \subcaption{}
 \end{minipage}
  \vspace{-2mm}
\caption{Centroid positions of whole fetuses' bones in a training set (a) without and (b) with BAC.}\label{fig:bac}
\end{figure}

\begin{figure}[b!]
 \begin{minipage}[b]{0.48\linewidth}
\centering   \includegraphics[width=\linewidth]{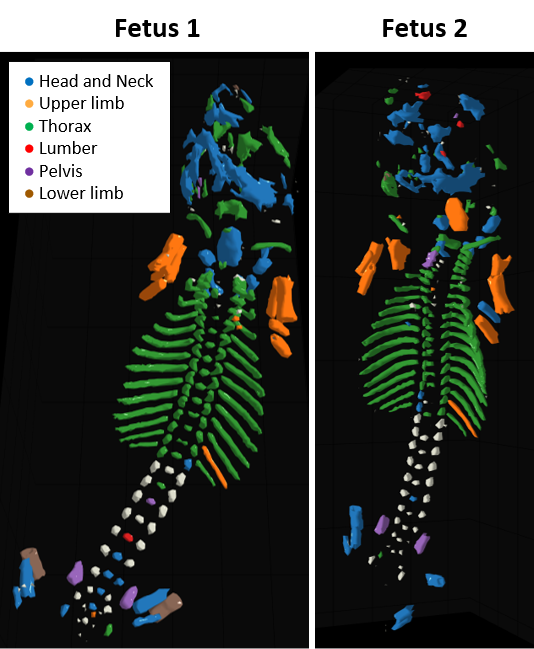}
  \subcaption{}
 \end{minipage}
 \begin{minipage}[b]{0.48\linewidth}
\centering   \includegraphics[width=\linewidth]{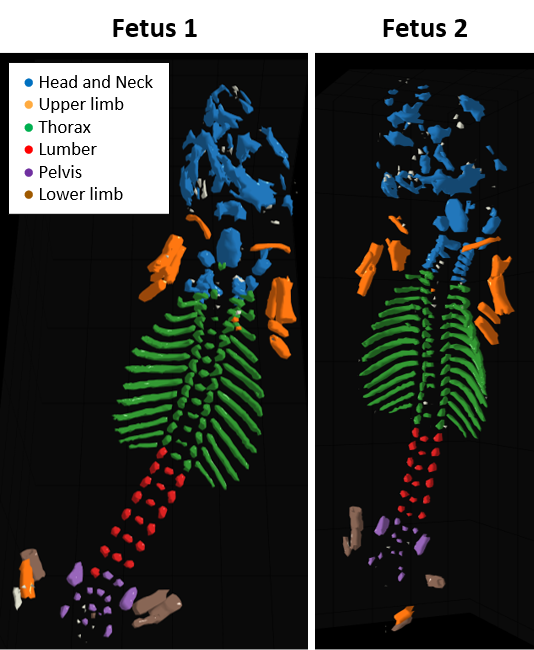}
  \subcaption{}
 \end{minipage}
 \vspace{-2mm}
\caption{Examples of skeletal labeling results obtained (a) without and (b) with BAC.}
\label{fig:res-postprocess}
\end{figure}

Table \ref{tab:res-cls} shows that introducing BAC improves skeletal labeling consistently, achieving accuracy/sensitivity improvement by $0.162$/$0.208$, respectively, for XGBoost. Training with both BAC and its spectral embedding always outperforms training with each feature alone by capturing both linear/non-linear relations. It obtains an accuracy of $0.900$ despite limited training data. The XGBoost and LightGBM perform similarly well, which implies that our feature engineering techniques can optimize various classifiers except for low-complex ones, such as SVM. As Fig. \ref{fig:bac} shows, BAC correctly aligns centroid positions of bones in terms of body parts. Because we input centroid positions to the classifiers, this could be a key factor in improving accuracy. Fig. \ref{fig:res-postprocess} visually reveals that the BAC mitigates misclassification in bones such as lower limb bones and neck/head bones. 


\begin{table}[b!]
\centering
\caption{Skeletal labeling results of XGBoost without/with skeletal relabeling(SR) on vertebral bodies/arches and ribs.}\label{tab:res-pp}
\begin{tabular}{ccccc}
\Hline\noalign{\smallskip} 
 \bfseries Bone & \bfseries SR & \bfseries Acc.[\%] & \bfseries Sen.[\%] & \bfseries Prc.[\%] \\\noalign{\smallskip}\hline\noalign{\smallskip}
 Vertebral body &  \xmark & 97.5 & 98.9 & 81.4\\
 Vertebral body &  \cmark & \textbf{99.7} & \textbf{99.5} & \textbf{97.8}\\
\noalign{\smallskip}\hline\noalign{\smallskip}
 Vertebral arch &  \xmark & 95.2 & \textbf{97.7} & 82.8\\
 Vertebral arch &  \cmark & \textbf{98.8} & 97.1 & \textbf{97.5}\\
\noalign{\smallskip}\hline\noalign{\smallskip}
 Rib &  \xmark & 99.2 & 97.3 & 97.6\\
 Rib &  \cmark & \textbf{99.3} & \textbf{97.7} & \textbf{97.9}\\
\noalign{\smallskip}\Hline\noalign{\smallskip}
\end{tabular}
\vspace{-1mm}
\end{table}

\newpage
\vspace{-2mm}
\subsection{Skeletal Abnormality Detection}\label{ssec:res-ad}
Table \ref{tab:res-pp} represents skeletal labeling results of XGBoost on vertebral bodies, vertebral arches, and ribs, where the skeletal abnormality is frequently observed. Introducing curve fitting-based skeletal relabeling leads to both high sensitivity/precision, effectively eliminating false positives often appearing outside the curves. Table \ref{tab:res-ad} indicates that combining multiple rules can achieve higher skeletal abnormality detection sensitivity than each rule alone.
The results include 11 False Negatives (FNs) and 9 False Positives (FPs). The main trigger of the FNs (i.e., nine FNs) is misdetection on either too small ribs or ribs connected to adjacent bones. Another FN is a thoracic vertebral body with abnormal shape while exhibiting a similar number of voxels to neighboring normal vertebral bodies. The other FN is a sternum, which is out of the scope of the rules. Meanwhile, regarding the FPs, Rule 1 causes eight FPs because of a shape discrepancy between stained specimens and their micro-CT scans; Rule 2 causes three FPs because some small bones are undetected (Two FPs are also caused by Rule 1). 

\begin{table}[t!]
\centering
\caption{Skeletal abnormality detection results obtained by 3 different rules on vertebral bodies and ribs.}\label{tab:res-ad}
\begin{tabular}{cccc}
\Hline\noalign{\smallskip} 
 \bfseries Rule & \bfseries Accuracy[\%] & \bfseries Sensitivity[\%] & \bfseries Precision[\%] \\\noalign{\smallskip}\hline\noalign{\smallskip}
     1 & 51.4 & 38.0 & 79.4\\
     2 & 74.3 & 66.2 & 94.0\\
     3 & 47.6 & 22.5 & \textbf{100.0}\\
     1+2+3 & \textbf{81.0} & \textbf{84.5} & 87.0\\
\noalign{\smallskip}\Hline\noalign{\smallskip}
\end{tabular}
\vspace{-3mm}
\end{table}

\vspace{-2mm}
\section{CONCLUSIONS}\label{sec:discussion}
\vspace{-2mm}
Our proposed approach―that thoroughly covers automated skeletal localization/labeling/abnormality detection―reliably detects rat fetuses' skeletal abnormalities with minimum whole-body micro-CT data collection and annotation effort. This could lead to the fast and quantitative automated screening of skeletal abnormalities for drug safety assessment.  Our various feature engineering techniques, in particular BAC-based registration, can remarkably improve the skeletal labeling without rigorous subject positioning in micro-CT.

In future work,  improve the skeletal localization and labeling to mitigate the misdetection of small bones and connected bones. Whereas we can reliably detect vertebral bodies and ribs with an accuracy of $0.810$, we plan to modify and add more detection rules to cover other bones and small ribs. Lastly, we might investigate fetuses of other small animals, such as rabbits.

\vspace{-2mm}
\section{ACKNOWLEDGEMENTS}\label{sec:discussion}
\vspace{-2mm}
The authors would like to thank Y. Sakai, Y. Aoki, and M. Fujiwara at Astellas Pharma Inc. for providing the dataset.






\bibliographystyle{IEEEbib}
\bibliography{refs}

\begin{thebibliography}{10}

\bibitem{winkelmann2009high}
Christopher~T Winkelmann and L~David Wise,
\newblock ``High-throughput micro-computed tomography imaging as a method to
  evaluate rat and rabbit fetal skeletal abnormalities for developmental
  toxicity studies,''
\newblock {\em J. Pharmacol. Toxicol. Methods}, vol. 59, no. 3, pp. 156--165,
  2009.

\bibitem{wise2010micro}
L~David Wise, Dahai Xue, and Christopher~T Winkelmann,
\newblock ``Micro-computed tomographic evaluation of fetal skeletal changes
  induced by all-trans-retinoic acid in rats and rabbits,''
\newblock {\em Birth Defects Res.}, vol. 89, no. 5, pp. 408--417, 2010.

\bibitem{french2010use}
Julian French, Neill Gingles, Jane Stewart, and Neil Woodhouse,
\newblock ``Use of magnetic resonance imaging ({MRI}) and micro-computed
  tomography (micro-{CT}) in the morphological examination of rat and rabbit
  fetuses from embryo-fetal development studies,''
\newblock {\em Reprod. Toxicol.}, vol. 30, no. 2, pp. 292--300, 2010.

\bibitem{dogdas2015characterization}
B.~Dogdas, A.~Chen, S.~Mehta, et~al.,
\newblock ``Characterization of bone abnormalities from micro-ct images for
  evaluating drug toxicity in developmental and reproductive toxicology
  ({DART}) studies,''
\newblock in {\em Proc. International Symposium on Biomedical Imaging (ISBI)},
  2015, pp. 671--674.

\bibitem{noguchi2020bone}
Shunjiro Noguchi, Mizuho Nishio, Masahiro Yakami, et~al.,
\newblock ``Bone segmentation on whole-body ct using convolutional neural
  network with novel data augmentation techniques,''
\newblock {\em Comput. Biol. Med.}, p. 103767, 2020.

\bibitem{belal2019deep}
Sarah~Lindgren Belal, May Sadik, Reza Kaboteh, et~al.,
\newblock ``Deep learning for segmentation of 49 selected bones in {CT} scans:
  First step in automated {PET/CT}-based {3D} quantification of skeletal
  metastases,''
\newblock {\em Eur. J. Radiol.}, vol. 113, pp. 89--95, 2019.

\bibitem{li201918f}
Hebei Li, Chongrui Xu, Bowen Xin, et~al.,
\newblock ``18{F-FDG PET/CT} radiomic analysis with machine learning for
  identifying bone marrow involvement in the patients with suspected relapsed
  acute leukemia,''
\newblock {\em Theranostics}, vol. 9, no. 16, pp. 4730, 2019.

\bibitem{solomon2018use}
Howard~M Solomon, Stacia Murzyn, Joyce Rendemonti, et~al.,
\newblock ``The use of micro-{CT} imaging to examine and illustrate fetal
  skeletal abnormalities in {Dutch} {Belted} rabbits and to prove concordance
  with alizarin red stained skeletal examination,''
\newblock {\em Birth Defects Res.}, vol. 110, no. 3, pp. 276--298, 2018.

\bibitem{chen2019localization}
Antong Chen, Dahai Xue, Tosha Shah, et~al.,
\newblock ``{Localization and labeling of cervical vertebral bones in the
  micro-{CT} images of rabbit fetuses using a {3D} deep convolutional neural
  network},''
\newblock in {\em Proc. SPIE Medical Imaging}, 2019, vol. 10949, p. 1094913.

\bibitem{viero1994three}
Timo Viero, Kai Oistamo, and Yrj{\"o} Neuvo,
\newblock ``Three-dimensional median-related filters for color image sequence
  filtering,''
\newblock {\em IEEE Trans. Circuits Syst. Video Technol.}, vol. 4, no. 2, pp.
  129--142, 1994.

\bibitem{watkins2011drug}
PB~Watkins,
\newblock ``Drug safety sciences and the bottleneck in drug development,''
\newblock {\em Clin. Pharm.}, vol. 89, no. 6, pp. 788--790, 2011.

\bibitem{shojaii2005automatic}
Rushin Shojaii, Javad Alirezaie, and Paul Babyn,
\newblock ``Automatic lung segmentation in {CT} images using watershed
  transform,''
\newblock in {\em Proc. IEEE International Conference on Image Processing
  (ICIP)}. IEEE, 2005, vol.~2, pp. II--1270.

\bibitem{lewiner2003efficient}
Thomas Lewiner, H{\'e}lio Lopes, Ant{\^o}nio~Wilson Vieira, and Geovan Tavares,
\newblock ``Efficient implementation of marching cubes' cases with topological
  guarantees,''
\newblock {\em J. Graphics Tools}, vol. 8, no. 2, pp. 1--15, 2003.

\bibitem{wold1987principal}
Svante Wold, Kim Esbensen, and Paul Geladi,
\newblock ``Principal component analysis,''
\newblock {\em Chemom. Intell. Lab. Syst.}, vol. 2, no. 1-3, pp. 37--52, 1987.

\bibitem{luo2003spectral}
Bin Luo, Richard~C Wilson, and Edwin~R Hancock,
\newblock ``Spectral embedding of graphs,''
\newblock {\em Pattern recognit.}, vol. 36, no. 10, pp. 2213--2230, 2003.

\bibitem{chen2015xgboost}
Tianqi Chen, Tong He, Michael Benesty, et~al.,
\newblock ``Xgboost: extreme gradient boosting,''
\newblock {\em R package version 0.4-2}, pp. 1--4, 2015.

\bibitem{noble2006support}
William~S Noble,
\newblock ``What is a support vector machine?,''
\newblock {\em Nat. Biotechnol.}, vol. 24, no. 12, pp. 1565--1567, 2006.

\bibitem{ke2017lightgbm}
Guolin Ke, Qi~Meng, Thomas Finley, et~al.,
\newblock ``Lightgbm: A highly efficient gradient boosting decision tree,''
\newblock in {\em Advances in Neural Information Processing Systems (NIPS)},
  2017, pp. 3146--3154.

\end{thebibliography}

\end{document}